\documentclass[11pt]{article}
\usepackage{helvet, graphicx}
\usepackage{hyperref}
\usepackage[usenames]{color}
\usepackage{amsmath,amssymb,amsfonts}
\usepackage{bm}
\usepackage[margin=0.75in]{geometry}
\usepackage[T1]{fontenc}
\usepackage{soul}

\usepackage[matrix,frame,arrow]{xypic}
\usepackage[braket]{Qcircuit}

\usepackage[font=small]{caption} 
\usepackage{cite}
\usepackage{setspace}

\newcommand{\er}{Er$^{3+}$ }
\newcommand{\upg}{\uparrow}
\newcommand{\downg}{\downarrow}

\title{Parallel single-shot measurement and coherent control of solid-state spins below the diffraction limit}

\author{Songtao Chen$^*$, Mouktik Raha\thanks{These authors contributed equally to this work}~, Christopher M.~Phenicie, Salim Ourari, Jeff D.~Thompson\thanks{jdthompson@princeton.edu}}
\date{%
    \centering
    \textit{Department of Electrical Engineering, Princeton University, Princeton, NJ 08544, USA}\\[2ex]%
    \today
}

\begin{document}

\maketitle

\begin{abstract}
Solid-state spin defects are a promising platform for quantum science and technology, having realized demonstrations of a variety of key components for quantum information processing, particularly in the area of quantum networks. An outstanding challenge for building larger-scale quantum systems with solid-state defects is realizing high-fidelity control over multiple defects with nanoscale separations, which is required to realize strong spin-spin interactions for multi-qubit logic and the creation of entangled states. In this work, we experimentally demonstrate an optical frequency-domain multiplexing technique, allowing high-fidelity initialization and single-shot spin measurement of six rare earth (Er$^{3+}$) ions, within the sub-wavelength volume of a single, silicon photonic crystal cavity. We also demonstrate sub-wavelength control over coherent spin rotations using an optical AC Stark shift. The demonstrated approach may be scaled to large numbers of ions with arbitrarily small separation, and is a significant step towards realizing strongly interacting atomic defect arrays with applications to quantum information processing and fundamental studies of many-body dynamics.
\end{abstract}

A central appeal of solid-state atomic defects for quantum technology is the possibility to realize strong dipolar interactions between closely spaced spins \cite{awschalom2013quantum}. This enables multi-qubit logic operations (to realize, for example, error correction \cite{taminiau2014universal,waldherr2014quantum} or deterministic teleportation over a quantum network \cite{pfaff2014unconditional}) as well as fundamental studies of many-body quantum phenomena \cite{choi2017observation}. Typically, these interactions are significant for defect separations less than several tens of nanometers. However, for optically-addressed spins, it is an open challenge to achieve simultaneous, high-fidelity initialization, control, and readout of spins separated by less than the diffraction limit of the addressing light, typically several hundred nanometers. Several techniques have been demonstrated to simultaneously address pairs of closely spaced NV centers, such as super-resolution microscopy \cite{maurer2010}, and variations in the Larmor frequency arising from different defect orientations \cite{neumann2010, dolde2013} or magnetic field gradients \cite{grinolds2011quantum}; however, these approaches have not been extended to high-fidelity operations such as single-shot spin readout, or to larger numbers of defects. Alternatively, an array of nuclear spins surrounding a single atomic defect can be distinguished by their positions in the gradient of the hyperfine coupling \cite{kolkowitz2012sensing, taminiau2012detection,zhao2012sensing}; while this approach has been used to generate entanglement between as many as 10 spins \cite{bradley2019}, it suffers from the bottleneck that all operations are performed through a single, central electron spin.

Rare earth ions (REIs) in solid-state hosts are a promising platform for many applications because of their demonstrated long coherence times (for example, exceeding 6 hours for Eu$^{3+}$ \cite{zhong2015optically}) as well as operation in the telecom band and compatibility with silicon photonics (in the case of Er$^{3+}$ \cite{dibos2018}). Furthermore, their unique spectral characteristics enable frequency-domain addressing of many defects within the same spatial volume. REIs experience random, static shifts of their optical transition frequencies that give rise to an inhomogeneous (ensemble) linewidth $\Gamma_\text{inh}$ (typically 1-10 GHz in crystalline hosts \cite{liu2006spectroscopic}) that is much broader than the homogeneous linewidth of an individual ion $\Gamma_\text{h}$ (typically < 1 MHz). In a given sample volume, this allows a large number of distinct subsets of ions to be separately addressed, of order $N_\text{ad} \approx \Gamma_\text{inh}/\Gamma_\text{h} > 10^3$. This approach can be applied to any solid-state emitter, in principle, but the uniquely small magnitude of $\Gamma_\text{inh}$ and $\Gamma_\text{h}$ in REIs allows the entire inhomogeneous distribution to be addressed with electro-optic sidebands on a single laser. Using spectral hole burning, this effect has been exploited in rare earth ensembles to realize multimode atomic memories for quantum networks \cite{de2008, saglamyurek2011}. Quantum gate architectures based on ensemble spectral-hole qubits have also been proposed \cite{ichimura2001, ohlsson2002, wesenberg2007, ahlefeldt2020} and demonstrated \cite{pryde2000solid, longdell2004PRA, longdell2004PRL}.

Frequency-domain addressing can also be used to address individual REIs within a diffraction-limited volume, if the total number of ions $N$ is less than $N_\text{ad}$. While detecting individual REIs is challenging owing to their low photon emission rates \cite{siyushev2014coherent,utikal2014spectroscopic,nakamura2014spectroscopy}, this problem can be overcome using Purcell enhancement in nanophotonic optical cavities \cite{dibos2018, zhong2018optically}, as exemplified by recent demonstrations of single-shot spin readout of single REIs \cite{raha2020, kindem2020}. In this work, we combine frequency-domain addressing and high fidelity optical control to realize initialization and single-shot spin readout of six \er spins with sub-micron separations, coupled to a single photonic crystal cavity. Additionally, we demonstrate individual spin control using an ion-selective AC Stark shift \cite{buckley2010spin}. Combined with existing techniques to create dense defect ensembles using ion implantation, this work is a significant step towards the creation of strongly-interacting defect arrays with single-particle control.

\begin{figure}[ht]
	\centering
    \includegraphics[width=86 mm]{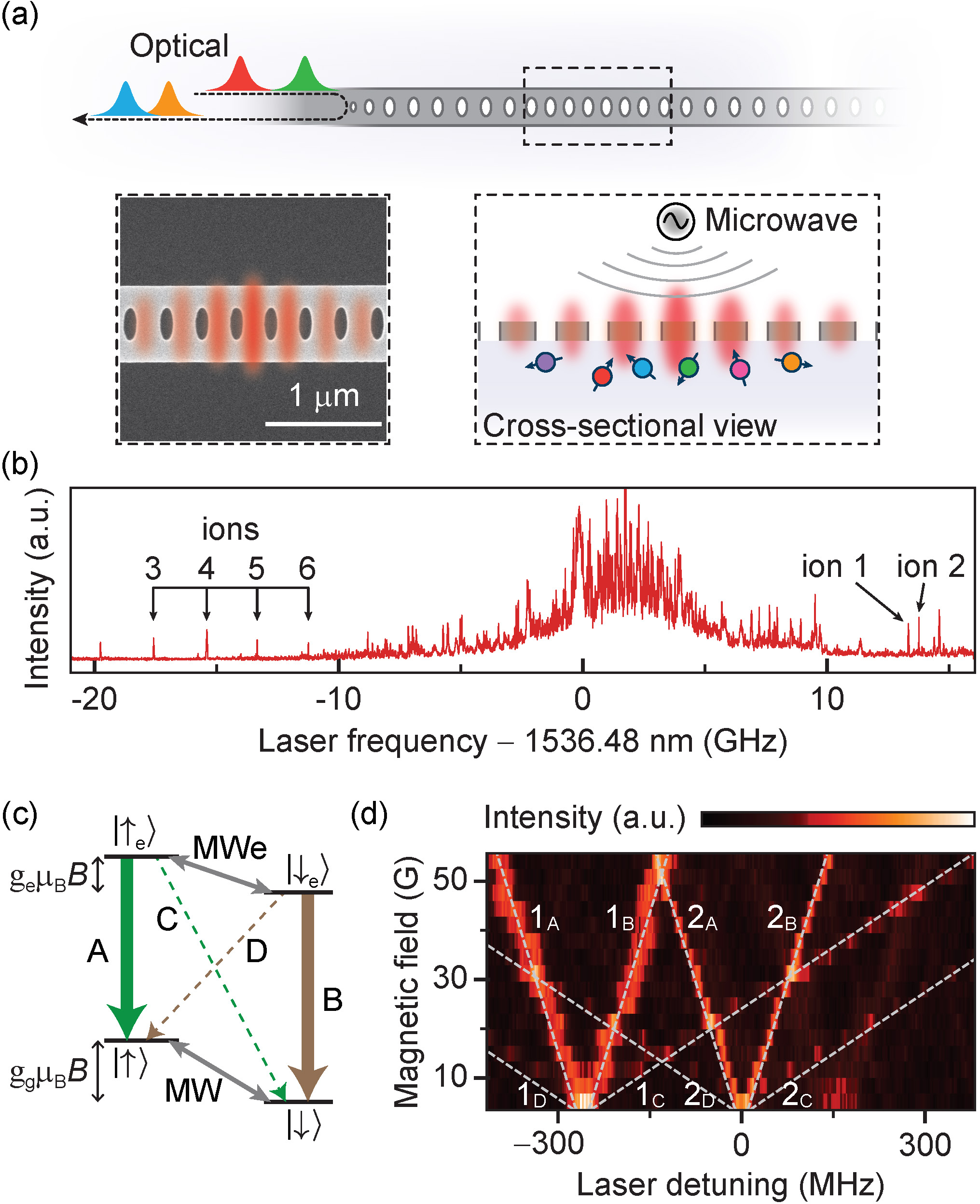}
    \caption{\textbf{Spectrally addressing multiple ions in a diffraction-limited volume.}
    \textbf{(a)} Schematic drawing of the device. (\emph{inset}) Scanning electron microscope image of a representative cavity, showing the extent of the optical mode.
    \textbf{(b)} PLE spectrum of \er ions in a single device with magnetic field $B=0$. Arrows indicate the six ions used in this work.
    \textbf{(c)} Level structure of \mbox{\er\!\!:YSO} in a magnetic field, with optical (A-D) and microwave (MW, MWe) transitions indicated.
    \textbf{(d)} PLE spectrum of ion~1 and ion~2 in the presence of a magnetic field (oriented along the $D_2$ axis of the YSO crystal). Zero detuning in this panel and subsequent figures refers to the ion~2 resonance when $B=0$. 
    }
	\label{fig:Figure1}
\end{figure}

Our experimental approach, depicted schematically in Fig.~\ref{fig:Figure1}a, consists of an Er$^{3+}$-doped Y$_2$SiO$_5$ (YSO) crystal coupled to a silicon photonic crystal cavity. The cavity enhances the emission rate of the ions \cite{dibos2018}, and modifies the selection rules to make the optical transitions highly cyclic, enabling single-shot spin readout \cite{raha2020}. The zero-field photoluminescence excitation (PLE) spectrum (Fig.~\ref{fig:Figure1}b) shows several hundred ions within the $0.05\,\mu$m$^3$ mode volume of the optical cavity, with an inhomogeneous linewidth of several GHz. We first focus on a pair of ions located in the blue tail of the inhomogeneous distribution, labeled ion~1 and ion~2, which couple to the cavity with Purcell factors of 330 and 200, respectively, when resonant with the cavity. Since the ions are addressed through a single-mode cavity, the optical signal provides no spatial information about the ions: they are within a single, diffraction-limited volume. Instead, the ions are addressed in the frequency domain, relying on a separation between their transitions of approximately 250 MHz, which is considerably larger than their linewidths (24 and 10 MHz) but smaller than the cavity linewidth of 4.2 GHz. In a magnetic field, each ion's optical transition splits into four lines that can be used to interface with its spin (Fig.~\ref{fig:Figure1}c,d).

\begin{figure}[ht]
	\centering
    \includegraphics[width=\linewidth]{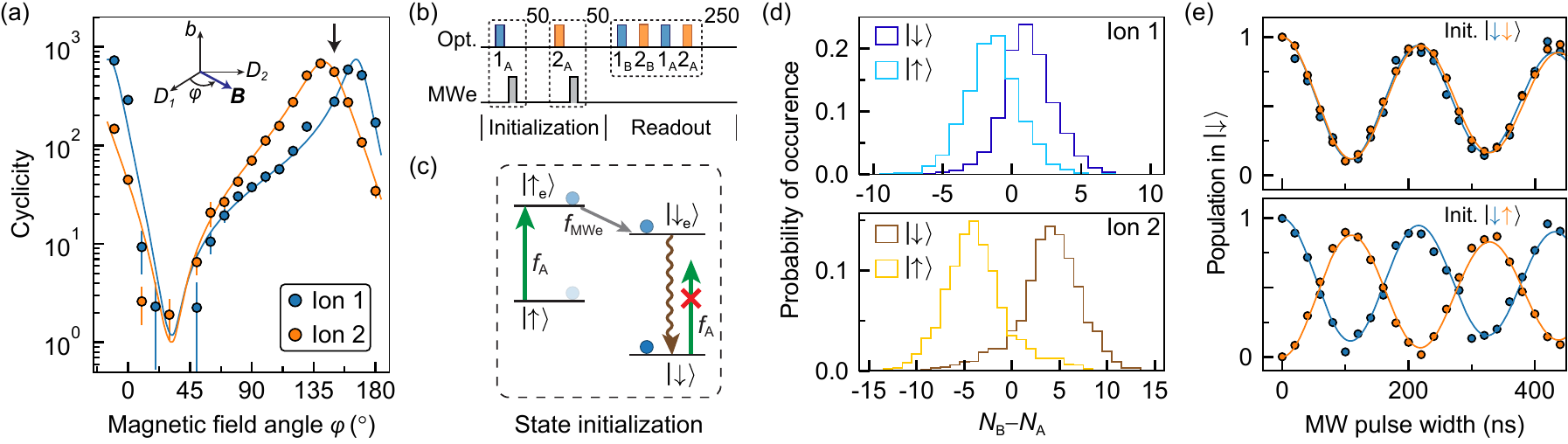}
    \caption{\textbf{Simultaneous initialization and readout of ions 1 and 2.}
    \textbf{(a)} Cyclicity of the optical transitions (defined as $1+\ \Gamma_\text{A,B}/\Gamma_\text{C,D}$, where $\Gamma_i$
    is the Purcell-enhanced decay rate on transition~$i$) as a function of magnetic field angle (see inset).
    The solid lines are fits to a theoretical model from Ref.~\cite{raha2020} and the black arrow indicates the orientation used in subsequent experiments ($\varphi=150^\circ$).
    \textbf{(b)} Pulse scheme used for spin initialization (in this case, to $\ket{\downg}$) and readout. All optical and microwave pulses are $\pi$ pulses.
    \textbf{(c)} Diagram of the initialization sequence. A dark state emerges in $\ket{\downg}$ from the combination of optical excitation of the spin-conserving A transition and microwave driving of the excited-state spin. Exciting B instead will initialize the spin to $\ket{\upg}$.
    \textbf{(d)} Photon histograms showing simultaneous single-shot readout for both ions. The ions are probed alternately on their A and B transitions, and $N_\text{B}-N_\text{A}$ denotes the difference in detected counts. $N_\text{B} > N_\text{A}$ indicates that the spin state is $\ket{\downg}$.
    \textbf{(e)} Simultaneous Rabi oscillations are observed while driving the ground state spin of both ions with a microwave pulse after initialization into the indicated states. The vertical axes is corrected for initialization and readout fidelity. 
    }
	\label{fig:Figure2}
\end{figure}

First, we demonstrate simultaneous initialization and single-shot spin measurement of ion~1 and ion~2. The measurement relies on cavity-enhanced cyclicity of the optical transitions, which is controlled by the alignment of the magnetic field to the local cavity polarization \cite{raha2020}. A magnetic field orientation of $\left[(\theta,\varphi)= (90,150)^\circ\right]$ allows high cyclicity for both ions, indicating similar cavity polarization at their respective positions. We initialize the spins by optical pumping, driving the excited state spin transition with microwaves to mix the spin levels (Fig.~\ref{fig:Figure2}b,c) \cite{afzelius2010, kindem2020}. Then, we perform a simultaneous single-shot spin measurement by alternately exciting the spin-conserving optical transitions (A,B) on each ion. For both initialization and measurement, the laser frequency is rapidly switched between transitions using a sideband from a fiber-coupled electro-optic modulator. We infer an initialization fidelity of $\geq$ 95\%, 97\% \cite{suppinfo} and an average readout fidelity of 76\%, 88\% (Fig.~\ref{fig:Figure2}d) for ion~1 and ion~2, respectively. The ions' spins can be coherently manipulated using microwave pulses that address both ions equally (Fig.~\ref{fig:Figure2}e), since the disorder in the Larmor frequency is much smaller than that of the optical transition. Details about the spin lifetime and coherence times can be found in the supplementary information \cite{suppinfo}.

\begin{figure}[ht]
	\centering
    \includegraphics[width=\linewidth]{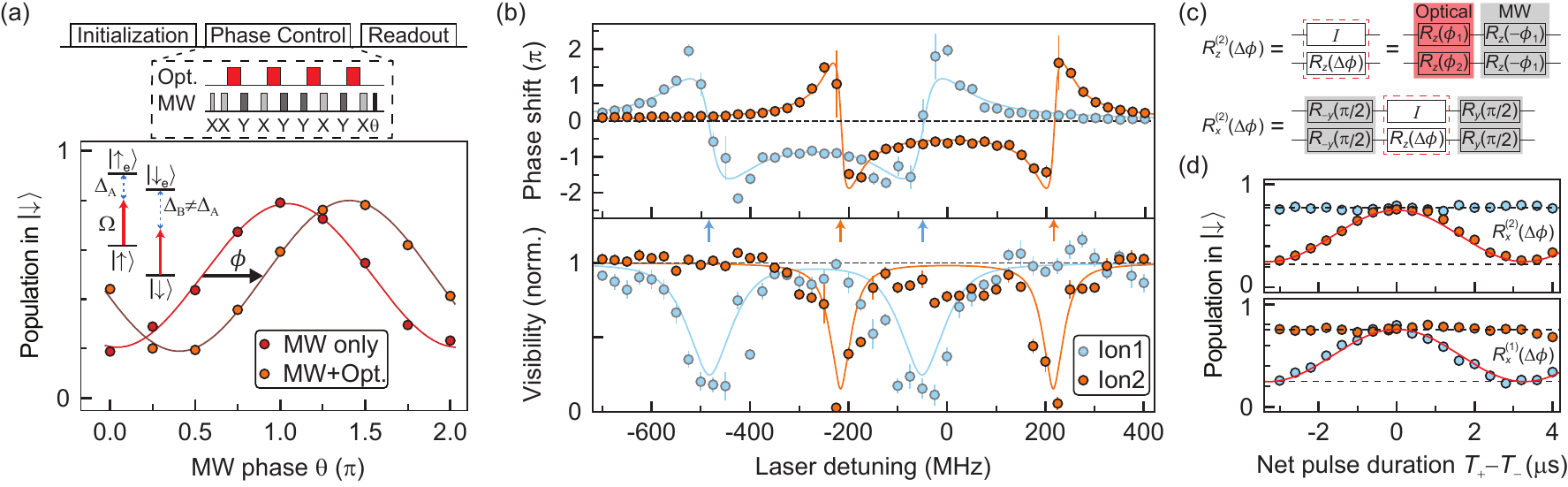}
    \caption{\textbf{Coherent optical spin rotation using the AC Stark shift. (a)} (\emph{upper}) Optical phase shifts are generated using a sequence of detuned optical pulses interleaved with a microwave-driven XY8 decoupling sequence. (\emph{lower}) The phase shift is detected by varying the phase of the final microwave $\pi/2$ pulse before measuring the spin population. The phase shift ($\phi$) and visibility are extracted from a sinusoidal fit.
    \textbf{(b)} Frequency dependence of the phase shift and change in visibility for each ion after a 2 $\mu$s optical pulse. The solid lines are numerical simulations including spectral diffusion. The arrows indicate the frequencies of the A,B transitions for each ion.
    \textbf{(c)} Circuit diagram for implementing $R_{z,x}^{(i)}(\Delta\phi)$ rotations.
    \textbf{(d)} Ion-selective Rabi oscillations $R_x^{(i)}(\Delta\phi)$. Positive (negative) phase shifts are generated by placing optical pulses after the odd (even) numbered $\pi$ pulses in the XY8 sequence, with total duration $T_+$ ($T_-$) (the laser detuning is 275~MHz). The dashed lines show the loss of contrast from dephasing measured in the absence of any optical pulses -- excess dephasing from the optical pulses is not observable.}
	\label{fig:Figure3}
\end{figure}

Next, we turn to demonstrating individually addressed spin manipulations. To achieve this, we utilize the AC Stark shift from a detuned optical pulse to induce a net phase shift $\phi$ between \ket{\upg} and \ket{\downg} \cite{buckley2010spin}. For each ion, the accumulated phase shift is $\phi = T\Omega ^2\left(\Delta^{-1}_{\text{B}}-\Delta^{-1}_{\text{A}}\right)/4$, where $T$ is the pulse duration, $\Omega$ is the optical Rabi frequency and $\Delta_{\text{A}}$, $\Delta_{\text{B}}$ are the detunings of the laser from the spin-conserving transitions A, B (Fig.~\ref{fig:Figure3}a, inset). For a given laser frequency and intensity, the detuning and Rabi frequency are different for each ion, enabling local control of the phase shift. To control $N$ ions, $N-1$ laser frequencies are needed, since microwave rotations provide an additional control axis \cite{suppinfo}. Here, with $N=2$, we can control both ions independently using a single laser frequency. In addition to the phase shift, there is also a loss of coherence from photon scattering and fluctuations in the optical transition frequency (\emph{e.g.}, from spectral diffusion), which happens at a rate $\Gamma' \propto \Gamma \Omega^2 \left(\Delta^{-2}_{\text{A}} + \Delta^{-2}_{\text{B}}\right)$, where $\Gamma$ is the effective transition linewidth \cite{suppinfo}.

We measure the optically induced phase shift and decoherence using Ramsey spectroscopy (Fig.~\ref{fig:Figure3}a). To select the optimum operational point, we characterize the phase shift and decoherence as a function of laser frequency (Fig.~\ref{fig:Figure3}b). The results are in good agreement with a theoretical model, with the optimal ratio of phase shift to decoherence attained for large detunings. 

In combination with a global microwave rotation $R_z(-\phi_1)$, the differential phase $\Delta \phi$ gives rise to a net rotation on ion 2 alone: $R^{(2)}_z(\Delta \phi) = \mathbb{I} \otimes R_z(\Delta \phi)$ (Fig.~\ref{fig:Figure3}c). Similarly, a global microwave rotation $R_z(-\phi_2)$ generates a rotation on ion~1 alone. Here, $R_{\hat{n}}(\alpha)$ denotes a rotation by an angle $\alpha$ about axis $\hat{n}$. Universal control of a single qubit requires arbitrary angle rotations around two orthogonal axes. However, global microwave rotations can transform ion-selective optical $z$ rotations into rotations around an arbitrary axis \cite{suppinfo}. As an example, we demonstrate rotations about the $x$ axis, $R^{(i)}_x(\Delta \phi)$, where $i=1,2$ denotes the target ion (Fig.~\ref{fig:Figure3}c,d), realizing more than $2\pi$ rotation as the optical pulse duration is varied.

\begin{figure}[!t]
	\centering
    \includegraphics[width = 89 mm]{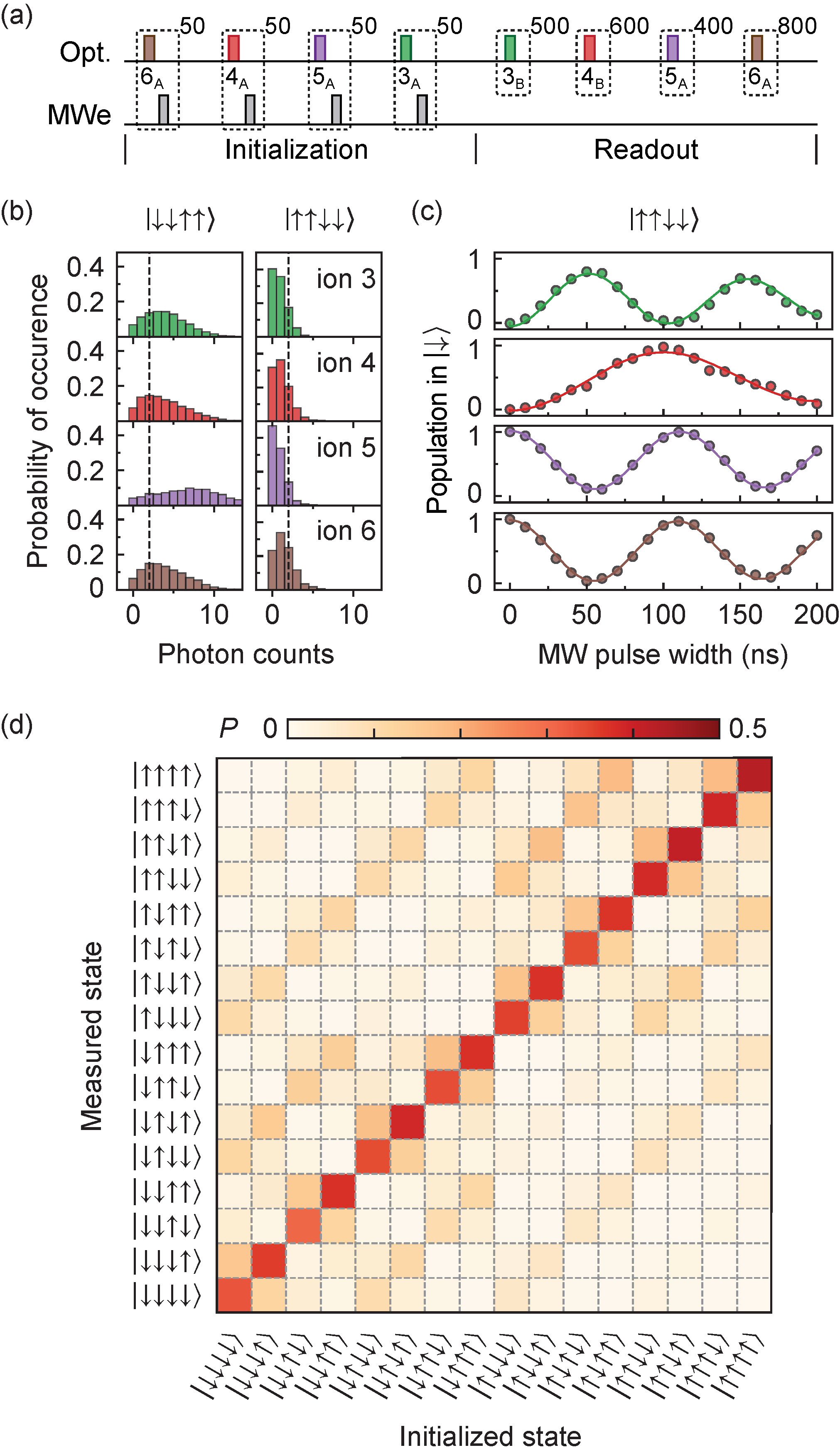}
    \caption{\textbf{Simultaneous initialization and readout of ions 3 -- 6.}
    \textbf{(a)} Pulse sequence for initialization and readout.
    \textbf{(b)} Histograms of detected photon counts for different initial states. The dashed lines indicate the threshold used for state discrimination. The spin state giving rise to higher counts depends on which transition is used for readout, which differs between ions.
    \textbf{(c)} Simultaneous microwave Rabi oscillations of all ions after being initialized to \ket{\upg\upg\downg\downg}. 
    \textbf{(d)} Single-shot readout results for all 16 four-ion initial states. The average probability to obtain the correct state for all four ions is $\bar{P} = 0.37$. 
    }\label{fig:Figure4}
\end{figure}

Lastly, we extend our approach to demonstrate simultaneous spin initialization and readout with four ions, labeled ion~3 through ion~6 (Fig.~1b). To access them, we shift the cavity resonance to $-14.8$~GHz (with respect to Fig.~1b), resulting in Purcell factors of 130, 260, 360, and 50. After choosing a magnetic field orientation that allows high cyclicity for all ions \cite{suppinfo}, we perform single-shot readout measurements. Because of the larger spread of these ions' frequencies (6.4 GHz) with respect to the cavity linewidth, it is advantageous to perform the readout using only one of the A or B transitions for each ion, whichever has larger Purcell enhancement (Fig.~\ref{fig:Figure4}a). The average readout fidelities for each ion are 80\%, 74\%, 87\%, and 71\%, respectively (Fig.~\ref{fig:Figure4}b). Although the ions are measured sequentially, the total measurement duration (300~ms) is much shorter than the ground state spin $T_1$ (typically > 10 s \cite{suppinfo}), such that the measurements are effectively simultaneous.

In Fig.~\ref{fig:Figure4}c, we show simultaneous microwave-driven Rabi oscillations on all four ions after initializing into $\ket{\upg \upg \downg \downg}$. Because ion 4 is situated in a rotated crystallographic site from the other ions, it has a different coupling to the microwave waveguide and correspondingly different Rabi frequency. In this measurement, the static field $B$ lies in the $D_1-D_2$ plane such that all ions have the same Larmor frequency, but we note that rotating $B$ out of this plane would make the ion 4 Larmor frequency different, enabling spectral addressing of its microwave transition. In Fig.~\ref{fig:Figure4}d, we show the initialization and single-shot measurement outcomes for all 16 four-spin states.

We have demonstrated simultaneous frequency-domain addressing of multiple \er spins within a diffraction-limited volume, realizing a complete set of operations: initialization, coherent control, and single-shot spin measurement. This approach may be extended in several ways. First, the total initialization and measurement time could be made independent of the number of ions by applying all tones simultaneously. In the case of readout, the simultaneous fluorescence from each ion could be discriminated using narrow bandwidth add-drop filters and separate detectors, or the measurement could be performed dispersively via the cavity reflection coefficient at the resonance frequency of each ion (assuming an improved atom-cavity cooperativity greater than 1 \cite{tiecke2014nanophotonic}). Second, the number of accessible ions is currently limited by spectral congestion in the center of the inhomogeneous distribution, but decreasing the total number of ions and $\Gamma_\text{h}$ should allow an entire ensemble of tens or hundreds of ions to be addressed, provided $\Gamma_\text{inh} \lesssim \kappa$. Lastly, the ions in this device have an average separation of 80 nm, where their dipolar interactions (1 kHz) are weaker than the decoherence from the $^{89}$Y nuclear spin bath in YSO. In future devices, implanting ions into small volumes \cite{bayn2015generation} in nuclear spin free hosts \cite{phenicie2019} will allow the creation of strongly interacting ensembles.

This work provides the first readily scalable path towards controlling dense defect arrays. This is immediately promising for frequency-multiplexed, telecom-wavelength quantum repeater nodes based on individual \er ions. Extensions to strongly interacting ensembles will enable more sophisticated quantum algorithms and fundamental studies of quantum many-body dynamics with single-particle control.

We gratefully acknowledge conversations with Nathalie de Leon and Mehmet Uysal. Support for this research was provided by the National Science Foundation (NSF, EFRI ACQUIRE program Grant No.~1640959), the Princeton Center for Complex Materials (PCCM), an NSF MRSEC (DMR-1420541), the Air Force Office of Scientific Research (Grant No.~FA9550-18-1-0081), the DARPA DRINQS program (Grant No.~D18AC00015). We acknowledge the use of Princeton's Imaging and Analysis Center, which is partially supported by PCCM, as well as the Princeton Micro-Nano Fabrication Lab and Quantum Device Nanofabrication Lab facilities. C.M.P.~was supported by the Department of Defense through the National Defense Science \& Engineering Graduate Fellowship Program.

\clearpage

\clearpage

\clearpage

\clearpage

\clearpage
\bibliography{phase_gate.bib}

\end{document}


\maketitle


\section{Experimental configuration}
\label{sec:setup}

\begin{figure}[!h]
	\centering
    \includegraphics[width=93.96 mm]{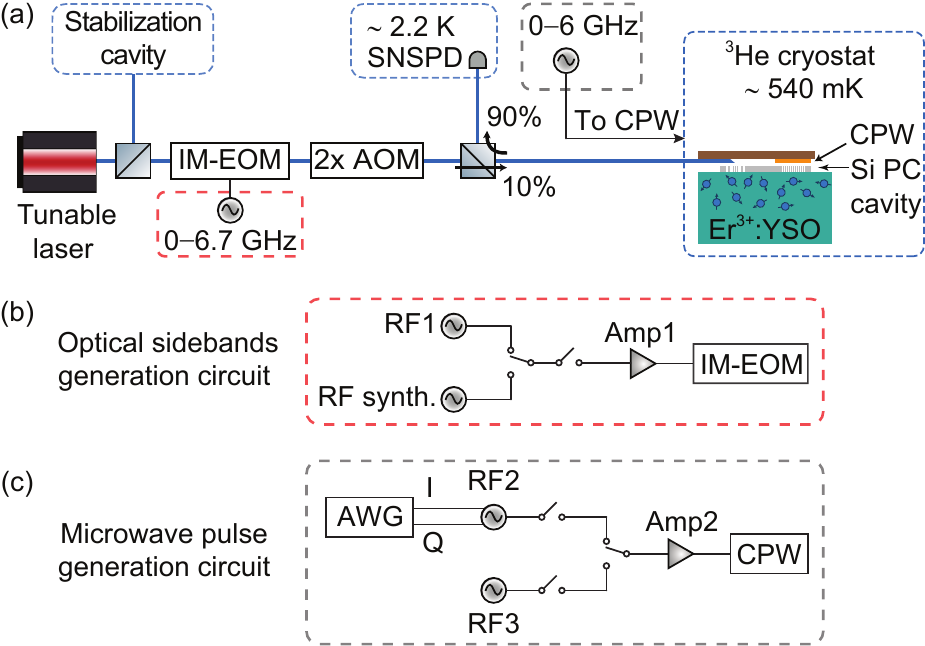}
    \caption{\textbf{Experimental setup.}
    \textbf{(a)} Sketch of the experimental setup.
    \textbf{(b,c)} Optical sidebands and microwave (MW) pulse generation circuit diagrams. See Sec.~\ref{sec:setup} for abbreviations.
    \textit{Additional component details}: Amp1, {Mini-Circuits} ZHL-4240W+; Amp2, {Mini-Circuits} ZHL-30W-252+; SPDT switches, {Mini-Circuits} ZASWA2-50DR-FA+.
    }\label{fig:setup_sketch} 
\end{figure}

\begin{figure}[!t]
	\centering
    \includegraphics[width=90.774 mm]{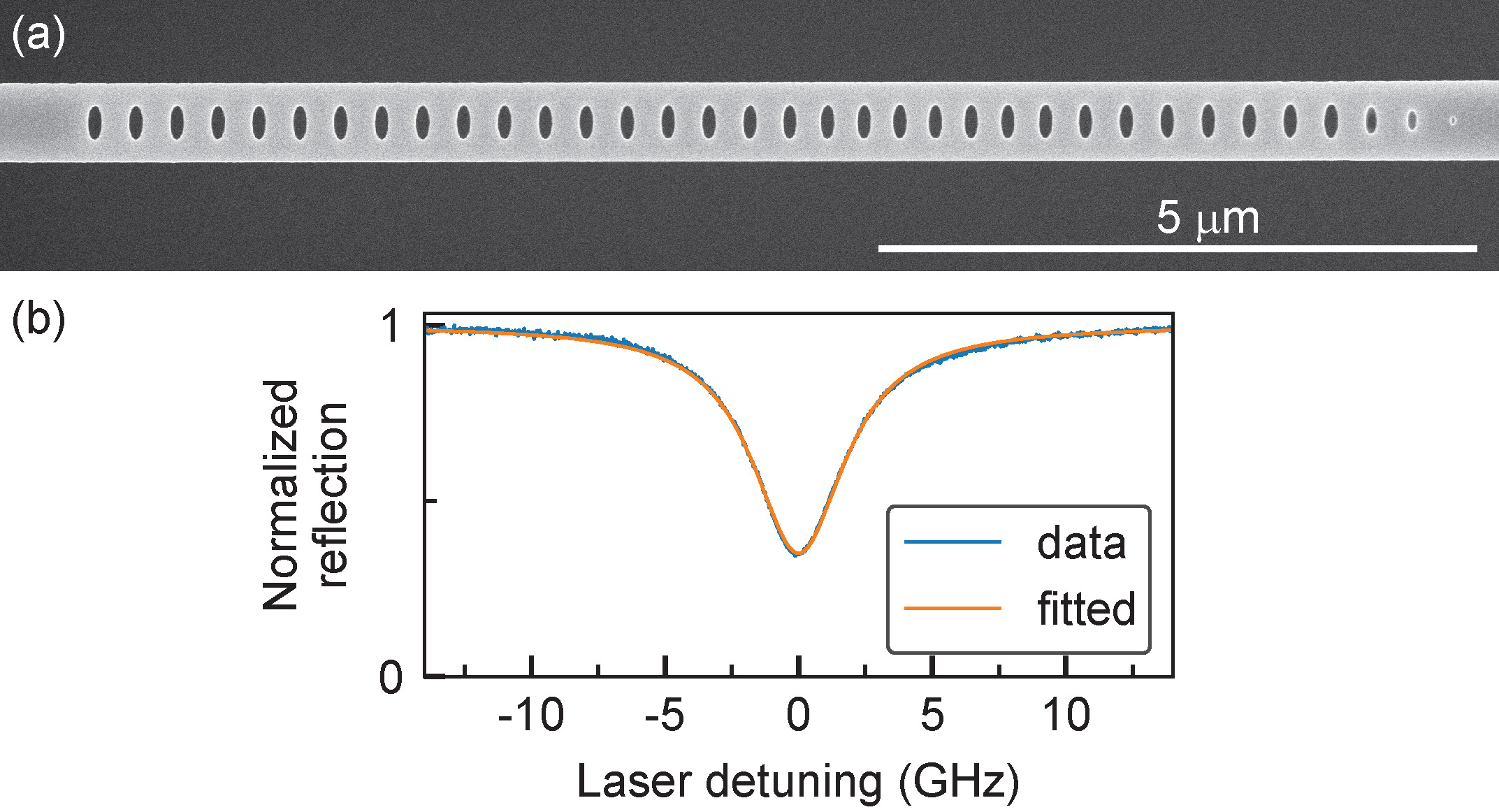}
    \caption{\textbf{PC cavity details.}
    \textbf{(a)} Scanning electron microscope image of a representative silicon photonic crystal cavity. \textbf{(b)} Reflection spectrum of the cavity used in this work with a measured quality factor $\text{Q} = 4.6 \times 10^4$.
    }\label{fig:cavityResonance}
\end{figure}

This section lays out a detailed description of our experimental setup (Fig.~\ref{fig:setup_sketch}a). A tunable laser (Toptica CTL1500), stabilized to a reference cavity (Stable Laser Systems), provides a single tone optical signal $f_\text{laser}$. Optical pulses are generated using an intensity modulating electro-optic modulator (IM-EOM, Lucent X-2623Y) and two acousto-optic modulators (AOMs, Isomet 1205C-1 and 1250C-868) in series. The two AOMs cumulatively provide a 142 dB optical power extinction ratio (one double-pass, one single-pass). The light enters a $^3$He cryostat (BlueFors LD250HE) in a single-mode optical fiber and is coupled to the PC cavity (Fig.~\ref{fig:cavityResonance}) using a grating coupler \cite{ding2013ultrahigh} with 40\% one-way efficiency. The fiber end, mounted on top of a three-axis nanopositioner (Attocube), is angle-polished to guide and couple light into the grating coupler \cite{li2014silicon}. Returning photons are separated using a 90:10 beamsplitter and detected in a superconducting nanowire single photon detector (SNSPD, Quantum Opus) in a separate, 2.2~K cryostat.

To enable fast switching of the laser frequency, we address the ions using RF sidebands generated by the IM-EOM (Fig.~\ref{fig:setup_sketch}b), with a DC bias chosen to suppress the carrier transmission. The initialization and readout sequences require switching the sideband frequency on the $\sim\!100\,\mu$s timescale, which is accomplished with a fast-switching RF generator (RF synth.,~Holzworth HSM6001B), while the optical phase shift pulse is generated by a separate synthesizer (RF1, SRS SG386). The spin transitions are driven with microwave magnetic fields generated by a coplanar waveguide attached to the end of the optical fiber. The ground and excited state spin transitions are driven by separate synthesizers (RF2 and RF3, respectively; SRS SG386) with signals being amplified to 21W before entering the cryostat (Fig.~\ref{fig:setup_sketch}c). An arbitrary waveform generator (AWG, Agilent 33622) drives an IQ mixer that modulates the signal from RF2 to generate MW pulses with different phases for the ground state.

\section{State-selective initialization of spins}
\label{sec:init}
\begin{figure}[!h]
	\centering
    \includegraphics[width = 155.19 mm]{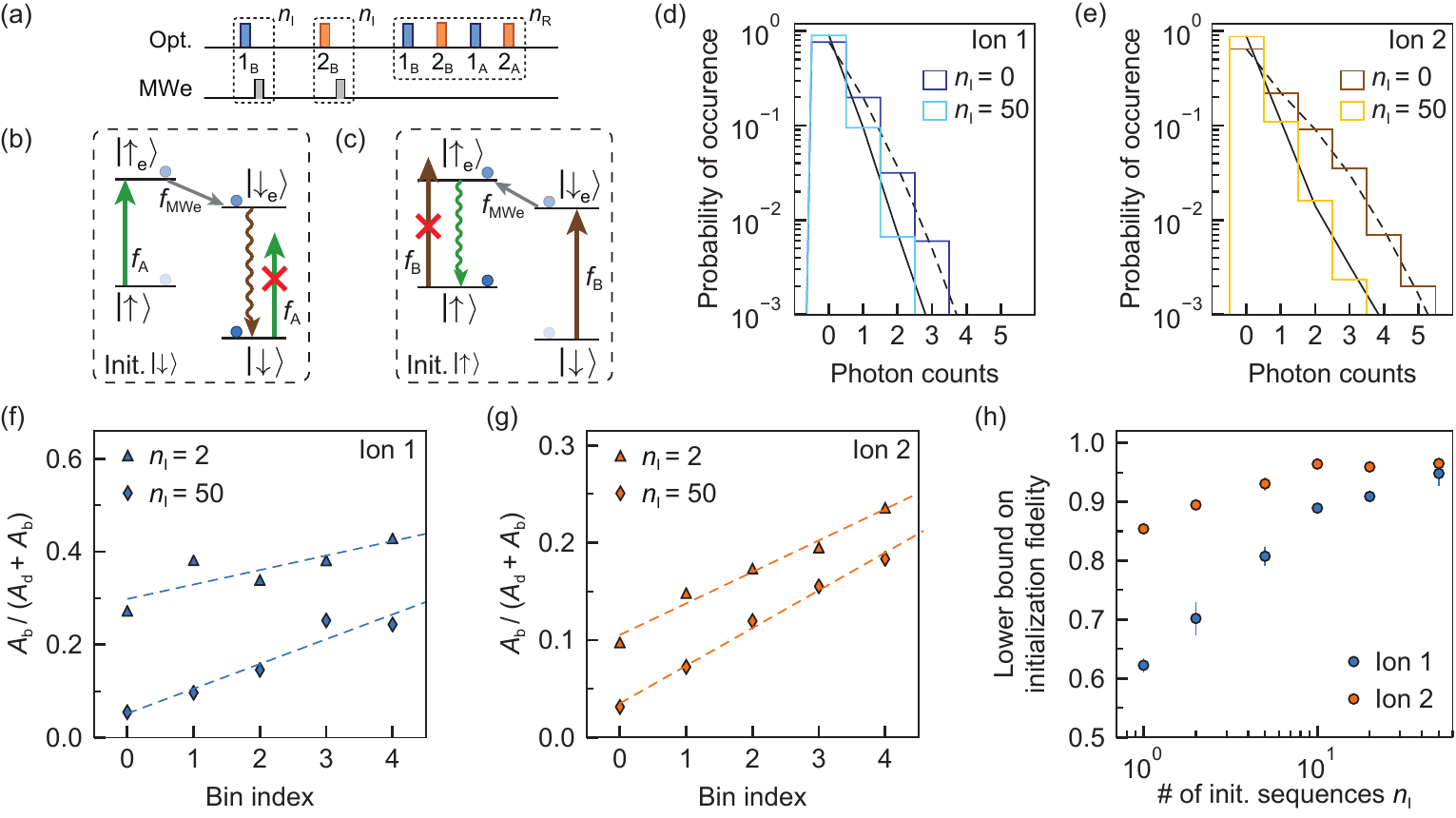}
    \caption{\textbf{Spin initialization of ions~1 and 2.}
    \textbf{(a)} Pulse scheme used for spin initialization (here, to $\ket{\upg}$) and readout.
    \textbf{(b,c)} Schematic diagrams depicting a combination of resonant optical and MW driving the excited state to initialize an \er ion to \ket{\downg} and \ket{\upg}.
    \textbf{(d,e)} Histograms of B photon counts when ion~1 and ion~2 are initialized to \ket{\upg} for different $n_\text{I}$. Black lines (solid and dashed) are fits to a bimodal Poisson distribution (see Sec.~\ref{sec:init}). Photons acquired during the first 50 readout pulses for each ion are considered here.
    \textbf{(f,g)} Fitted amplitudes of the bimodal distribution. Each bin consists of 50 consecutive readout pulses. Fitting the data to a linear function reveals \mbox{$y$-intercept} as an upper bound on the initialization \emph{infidelity}. 
    \textbf{(h)} Initialization fidelity for the two ions using varying number of initialization sequences $n_\text{I}$.
    }\label{fig:initialization}
\end{figure}

In this section, we provide a detailed characterization of the state-selective initialization protocol. A sketch of the protocol for initializing the spins is shown in Fig.~\ref{fig:initialization}b,c.
The initialization fidelity is significantly better than the readout fidelity, such that we cannot directly probe the initialization error. To get a better estimate of the initialization fidelity, we look at the photon number distributions during the measurement while implementing the pulse sequence in Fig.~\ref{fig:initialization}a.
In an ideal case of perfect initialization to \ket{\upg}, a histogram of detected photon counts while exciting the ion using transition B exhibits a dark count-limited Poisson distribution. Imperfect initialization leads to the presence of an additional Poisson distribution with a higher mean photon count due to residual population in \ket{\downg}.
Representative histograms of B photons while ion~1 and ion~2 are initialized to \ket{\upg} are shown in Fig.~\ref{fig:initialization}d,e. The data is fitted to a bimodal Poisson distribution of the form $A_\text{d} \exp(-\mu_\text{d}) \mu_\text{d}^x / x! + A_\text{b} \exp(-\mu_\text{b}) \mu_\text{b}^x / x!$, where 
$\mu_\text{b}, \mu_\text{d}$ are the mean photon counts while exciting each ion on the A,B transitions (when the ions are initialized to \ket{\upg}), and $A_\text{b}, A_\text{d}$ are the fitted amplitudes of corresponding distributions. If $\mu_\text{b}$ is big enough that the histogram is bimodal, $A_\text{b}$ can be fitted with high precision. 

However, finite cyclicity of the spin-conserving transitions A,B introduces a small probability ($\propto$ $n_\text{R}$, for $n_\text{R} \ll$ cyclicity) to flip the electron spin during the readout process. To that end, while the readout sequence involves $n_\text{R}=250$ pulses, we compute photon count histograms considering each group of 50 consecutive readout pulses (denoted as a ``bin'') at a time, and fit the histograms to the aforementioned bimodal Poisson distribution. We conservatively interpret $A_\text{b}/(A_\text{b}+A_\text{d})$ to be the probability that the spin was in the wrong state at the \emph{beginning} of the bin. As expected, the extracted probability value increases linearly with the bin index (Fig.~\ref{fig:initialization}f,g). By fitting a linear function to the data, we interpret the $y$-intercept to be an upper bound on the initialization \emph{infidelity}. We show the result of this fidelity analysis for varying number of repetitions of the initialization sequence, $n_\text{I}$, in Fig.~\ref{fig:initialization}h, revealing initialization fidelity for ion~1 and ion~2 to be at least 95\% and 97\% respectively, using $n_\text{I} = 50$. 
The lower bound is limited by the finite probability of optically pumping the ion inside the first bin, and we believe that the saturated initialization fidelity should be nearly perfect, limited only by off-resonant driving of the wrong transition. The slower rate of saturation and inferior initialization fidelity of ion~1 can be attributed to its larger spectral diffusion, leading to less efficient optical excitation probability per pulse.

\section{Spin coherence}
\label{sec:spin_dynamics}
\begin{figure}[!b]
	\centering
    \includegraphics[width=171.35 mm]{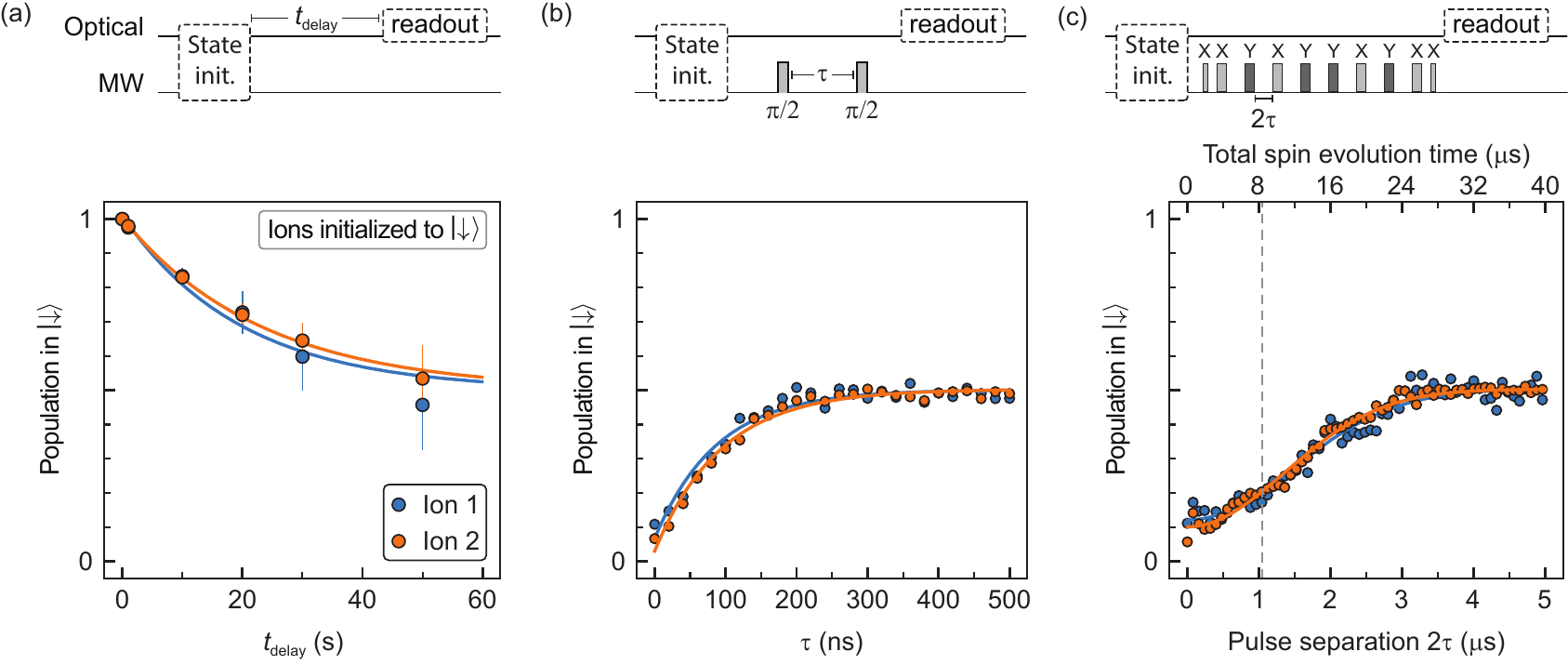}
    \caption{\textbf{Spin coherence of ion~1 and ion~2.}
    \textbf{(a)} Measurement of $T_1$. \textbf{(b)} Measurement of $T_2^*$. \textbf{(c)} Measurement of coherence during XY8 sequence. The dashed line denotes the pulse separation ($2\tau=1042$~ns) used for all experiments in Fig.~3. All measurements were performed with the magnetic field configuration $(B,\theta,\phi)= (112~\text{G},90^\circ,150^\circ)$.
    }\label{fig:spin_dynamics}
\end{figure}

We measure the ground state spin $T_1$ value by initializing each ion to the $\ket{\downg}$ state and performing spin readout at varying delay times ($t_\text{delay}$); a single exponential fit to the data gives $T_1$ of $19.9 \pm 4.1$~seconds and $23.3 \pm 1.3$~seconds for ion~1 and ion~2, respectively (Fig.~\ref{fig:spin_dynamics}a). Ramsey measurements reveal $T_2^*$ of $88 \pm 9$~ns and $94 \pm 6$~ns, respectively, extracted from fitting a single exponential to the data (Fig.~\ref{fig:spin_dynamics}b). For the optical phase control, we embed optical pulses within an XY8 sequence. The coherence decay during an XY8 sequence of total length $t$ is well-described by a Gaussian $\exp[-(t/T_2)^2]$ with $T_{2,\text{XY8}}$ of $16.5 \pm 0.5$~$\mu$s and $15.3 \pm 0.3$~$\mu$s.

We have not measured the spin coherence of ions 3 -- 6 in the course of this work, but note that the spin of ion~5 was partially characterized in Ref.~\cite{raha2020} at a slightly different magnetic field orientation (in that work, it was referred as ``ion 3''), with values of $T_1=45 \pm 4$~seconds, $T_2^* = 125 \pm 5$~ns, $T_\text{2,Hahn echo} = 3.3 \pm 0.2$~$\mu$s.

\section{Single qubit gates}
\subsection{Theory of ion-selective phase-shifts and loss of visiblity}
\label{sec:analyPhase}

We apply detuned optical pulses to accumulate phase shift using AC Stark effect. Consider the atom as a two-level system \{\ket{g},\ket{e}\} with a transition splitting of $\hbar\omega_0$, driven by a laser at frequency $\omega$. Under the dipole approximation, the effective non-Hermitian Hamiltonian in the rotating frame of the laser field (making the rotating-wave approximation) is given by:
\begin{equation}\tag{S1}
    H = \frac{\hbar\Omega}{2} (\ket{g}\bra{e} + \ket{e}\bra{g}) - \hbar(\Delta+i\Gamma/2)\ket{e}\bra{e},
\end{equation}
where $\Omega$ is the Rabi frequency, $\Delta = \omega-\omega_0$ is the detuning of the optical drive from the transition frequency, and $\Gamma$ is the spontaneous emission rate. The perturbed ground state and its corresponding energy shift are:
\begin{equation}\tag{S2}
    \ket{\psi} = \ket{g} + \frac{\sqrt{(\Delta+i\Gamma/2)^2+\Omega^2}-(\Delta+i\Gamma/2)}{\Omega} \ket{e} \approx \ket{g} + \frac{\Omega}{2(\Delta+i\Gamma/2)} \ket{e}
\end{equation}
\begin{equation}\tag{S3}
    \Delta E = \text{Re} \left\{ \frac{\hbar}{2}\left[\sqrt{(\Delta+i\Gamma/2)^2+\Omega^2}-(\Delta+i\Gamma/2)\right]\right\} \approx \hbar\frac{\Omega^2}{4}\frac{\Delta}{\Delta^2+\Gamma^2/4}
\end{equation}
for $\Omega \ll \sqrt{\Delta^2+\Gamma^2/4}$. For a given laser frequency, the two spin-conserving transitions A,B of an \er ion experience dissimilar detunings, denoted by $\Delta_\text{A},\Delta_\text{B}$. However, $\Omega$ must be the same for the two transitions since the spin states originating from the same doublet are related to each other via time-reversal symmetry \cite{raha2020}. Treating these transitions as two independent two-level systems, the net phase shift imprinted between $\ket{\downg}$ and $\ket{\upg}$ resulting from the detuned optical pulse is:
\begin{equation}\tag{S4}
    \phi = T \frac{\Omega^2}{4} \left(\frac{\Delta_\text{B}}{\Delta_\text{B}^2+\Gamma^2/4}-\frac{\Delta_\text{A}}{\Delta_\text{A}^2+\Gamma^2/4}\right),
\end{equation}
where $T$ is the optical pulse width. On the other hand, due to mixing of $\ket{e}$ in the perturbed ground state (Eq.~S2), there is a possibility of spontaneous emission from the excited state, which reduces the visibility ($\mathcal{V}$) of the fringes in the phase shift measurements. This can be calculated perturbatively from the excited state population $\Omega^2/(4\Delta^2+\Gamma^2)$ for each transition as:
\begin{equation}\tag{S5}
    \Delta \mathcal{V} = 1 - \exp \left[-  T \Gamma \frac{\Omega^2}{4} \frac{1}{2} \left(\frac{1}{\Delta_\text{A}^{2}+\Gamma^2/4} + \frac{1}{\Delta_\text{B}^{2}+\Gamma^2/4}\right) \right].
\end{equation}
For far-detuned optical drive pulses ($\Delta_\text{A},\Delta_\text{B} \gg \Gamma$), Eq.~S4 and S5 reduces to the following forms:
\begin{equation}\tag{S6}
    \phi = T \frac{\Omega^2}{4} \left(\frac{1}{\Delta_\text{B}}-\frac{1}{\Delta_\text{A}}\right),\quad\quad
    \Delta \mathcal{V} = 1 - \exp \left[-  T \Gamma \frac{\Omega^2}{4} \frac{1}{2} \left(\frac{1}{\Delta_\text{A}^{2}} + \frac{1}{\Delta_\text{B}^{2}}\right) \right].
\end{equation}

\subsection{Linewidth broadening}
\label{sec:linebroadening}

\begin{figure}[!t]
	\centering
    \hspace*{4cm}\includegraphics[width=121.294 mm]{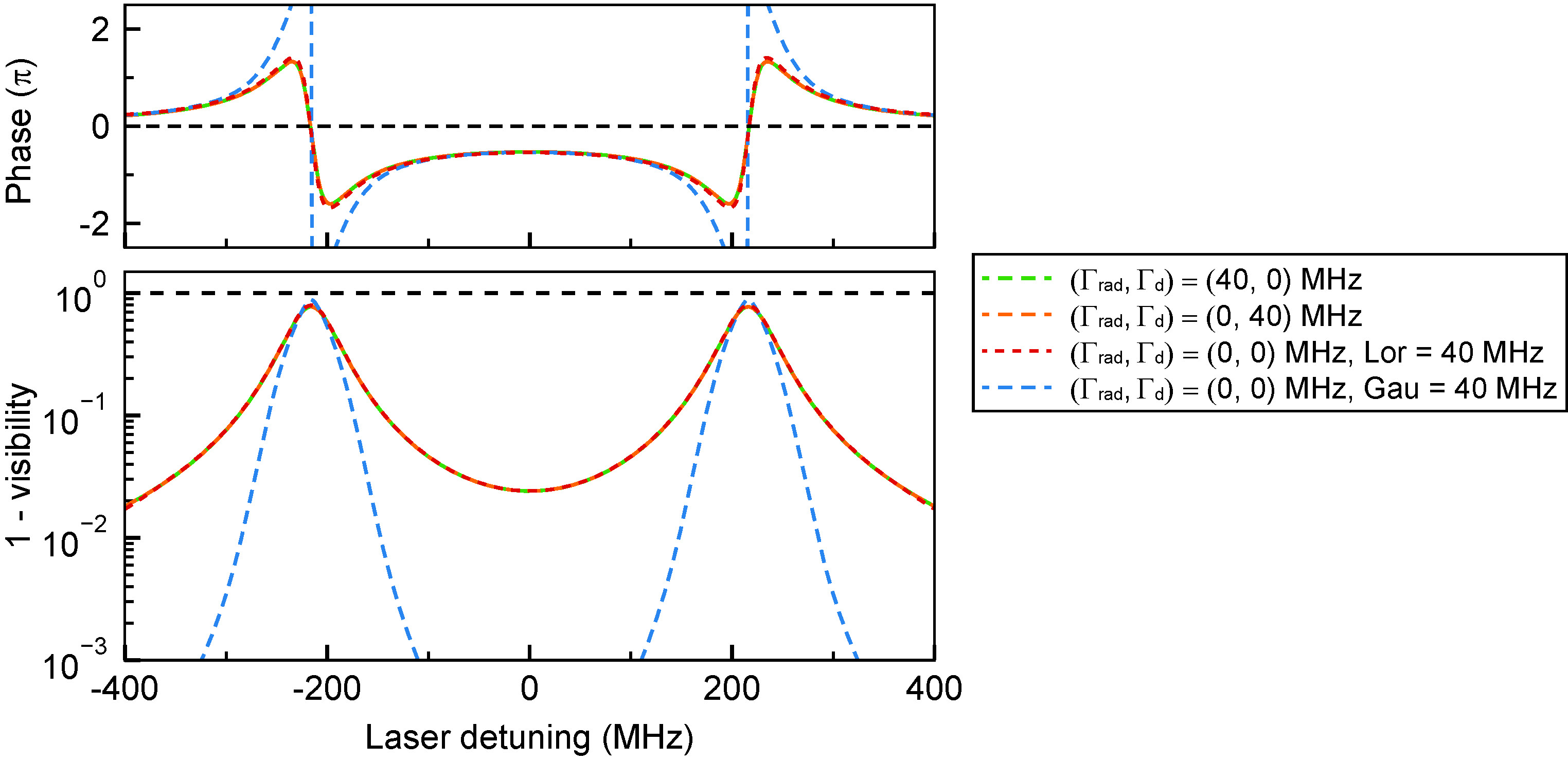}
    \caption{\textbf{Simulation results for the phase shift and change in visibility}. Numerical simulation of the master equation with line broadening from different sources. To consider the slow spectral diffusion effect, we average the results over different spectral profiles including Gaussian (``Gau.'') and Lorentzian (``Lor.''). $\Gamma_\text{rad}$ and $\Gamma_\text{d}$ represent broadening due to radiative relaxation and dephasing, respectively. The linewidths listed for Gaussian and Lorentzian profiles are FWHM.
    }\label{fig:qutip_linewidth}
\end{figure}

In our experiments, the observed linewidth is more than two orders of magnitude broader than the spontaneous emission rate, presumably because of spectral diffusion. While the calculation above estimates the loss of coherence from radiative decay of the excited state, the spin can also dephase from fluctuations in the optical transition frequency that cause uncertainty in the AC Stark shift. From numerical simulations of the master equation (Fig.~\ref{fig:qutip_linewidth}), we find that the functional form of the loss of coherence (Eq.~S5) is identical for the following cases: pure radiative decay, pure dephasing, or slow diffusion of the transition with a Lorentzian probability distribution. If multiple processes are present, the functional form is unchanged and their (Lorentzian) linewidths sum. Conversely, slow diffusion with a Gaussian probability distribution gives rise to a different line shape. In the experiment, we do not probe the loss or lineshape with sufficient precision to distinguish these cases.

\subsection{Simultaneous arbitrary rotations on two qubits}
\label{sec:unirot}
 
Universal control of a single qubit requires arbitrary rotations around two orthogonal axes. MW control alone imparts arbitrary but identical rotations on all spin qubits since it addresses the spins \emph{globally}. In this section, we theoretically show that arbitrary rotations on multiple spins can be achieved by combining MW rotations and ion-selective optical $z$ rotations.

We start with the simple case of two qubits. Arbitrary unitary operations on two qubits of the form $U \otimes U'$ can be decomposed into two consecutive operations $U \otimes \mathbb{I}$ and $\mathbb{I} \otimes U'$. First, let's implement the operation $U \otimes \mathbb{I}$. Consider the following quantum circuit:
\begin{equation}\tag{S7}
\Qcircuit @C=1.0em @R=.7em {
& \gate{V^{-1}} & \gate{R^\text{opt}_z(\phi_{1})} & \gate{V} & \qw \\
& \gate{V^{-1}} & \gate{R^\text{opt}_z(\phi_{2})} & \gate{V} & \qw \\
}
\end{equation}
where $R^\text{opt}_z(\phi_j)$ is the ion-selective optical $z$ rotation and $V$ is a global unitary rotation using MW pulses. To implement identity operation on qubit 2, we can insert a global $z$ rotation $R_z(-\phi_2)$, which can be accomplished by shifting the phase of MW pulses in $V$ relative to $V^{-1}$:
\begin{equation}\tag{S8}
\Qcircuit @C=1.0em @R=.7em {
& \gate{V^{-1}} & \gate{R^\text{opt}_z(\phi_{1})} & \gate{R_z(-\phi_2)} & \gate{V} & \qw \\
& \gate{V^{-1}} & \gate{R^\text{opt}_z(\phi_{2})} & \gate{R_z(-\phi_2)} & \gate{V} & \qw \\
}
\end{equation}
Since the phase shifts on qubit 2 cancel each other, we have $VV^{-1} = \mathbb{I}$. Therefore, the problem statement of implementing $U \otimes \mathbb{I}$ reduces to finding a matrix V such that $VR_z(\phi)V^{-1}=U$, where $\phi = \phi_1 - \phi_2$. Without loss of generality, $U$ can be represented as a rotation of angle $\alpha$ about an arbitrary axis $\hat{n}$: $U=R_{\hat{n}}(\alpha) \equiv \exp(-i\alpha \, \hat{n}\cdot\vec{\sigma}/2) =\cos(\alpha/2) \mathbb{I} -i \sin(\alpha/2)\hat{n}\cdot\vec{\sigma}$, where $\vec{\sigma} = (\sigma_x,\sigma_y,\sigma_z)$ is the Pauli vector. Denoting the unit vector $\hat{n}$ as $\hat{n}=(\sin{\theta}\cos{\beta},\sin{\theta}\sin{\beta},\cos\theta)$, we can write $U$ in the matrix notation as,
\begin{equation}\tag{S9}
    U = 
    \begin{bmatrix}
        \cos(\alpha/2)-i\sin(\alpha/2)\cos\theta & -ie^{-i\beta}\sin(\alpha/2)\sin{\theta}\\
        -ie^{i\beta}\sin(\alpha/2)\sin{\theta} & \cos(\alpha/2)+i\sin(\alpha/2)\cos\theta
    \end{bmatrix}
\end{equation}
The matrix $U$ can be diagonalized, revealing eigenvalues of $e^{\pm i\alpha/2}$. By using the corresponding eigenvectors ($u_1$, $u_2$) we can construct matrix $V$ as,
\begin{equation}\tag{S10}
    V = [u_1,u_2]=
    \begin{bmatrix}
        \cos(\theta/2) & -e^{-i\beta}\sin(\theta/2)\\
        e^{i\beta}\sin(\theta/2) & \cos(\theta/2)
    \end{bmatrix}
\end{equation}
so that $V^{-1}UV$ is diagonalized, leading to,
\begin{equation}\tag{S11}
    V^{-1}UV = 
    \begin{bmatrix}
        e^{-i\alpha/2} & 0\\
        0 & e^{i\alpha/2}
    \end{bmatrix}
    = R_z(\alpha)
\end{equation}
Therefore, the constructed matrix $V$ satisfies the equation $VR_z(\phi)V^{-1}=U$ if $\phi=\alpha$, thereby implementing $U \otimes \mathbb{I}$. Moreover, the matrix $V$ can also be expressed in terms of rotations about $y$ and $z$ axes, as $V=R_z(\beta)R_y(\theta)R_z(-\beta)$, which makes it easy to implement using MW pulses. $\mathbb{I} \otimes U'$ can also be implemented using the same sequence (S8) but with $R_z(-\phi_1)$ instead. 

As an example, for arbitrary $x$ rotations ($\theta = 90^{\circ},~\beta = 0^{\circ}$), $V=R_y(\pi/2)$, which has been utilized to perform arbitrary $x$ rotations in Fig.~3c,d of the main text. This protocol could be extended to $N > 2$ ions by considering $N-1$ ion-selective optical $z$ rotations together with global MW control.

\section{Additional measurements on four ions}
\label{sec:fourIons}

In this section, we provide additional details of measurements performed on the four ions labeled ion~3 through ion~6 (Fig.~1b). The PC cavity is tuned to a spectral location such that spectral lines of all the four ions are situated within a cavity linewidth away from the cavity resonance.

\begin{figure}[!h]
	\centering
    \includegraphics[width = 50.3 mm]{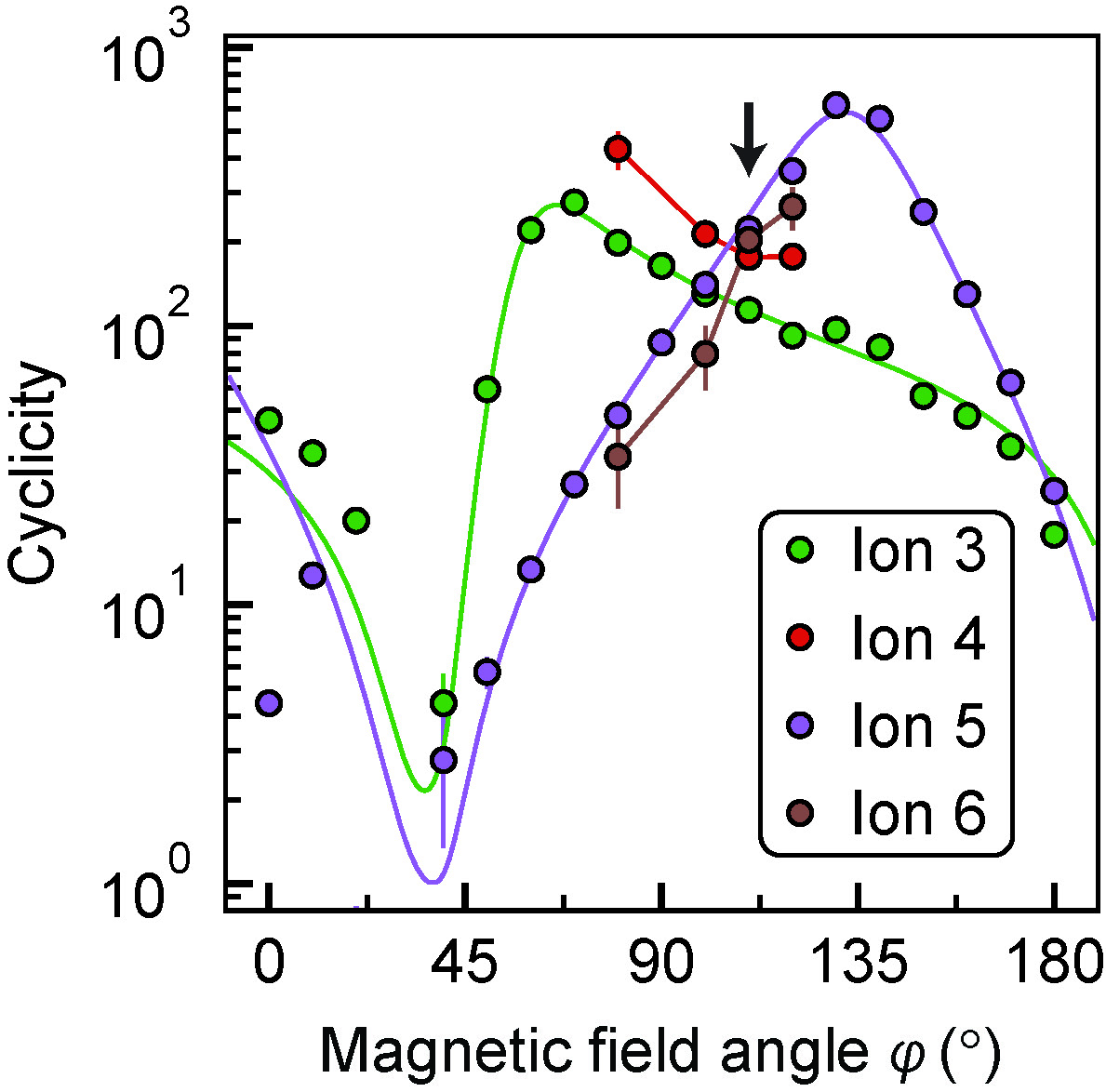}
    \caption{
    \textbf{Four-ion cyclicity.} Magnetic field orientation dependence of cyclicity of the optical transitions measured for four ions ($B$ = 112 G, $\theta = 90^\circ$). Results for ion~3 and ion~5 are fitted to a theoretical model from Ref.~\cite{raha2020}. Black arrow at $\varphi = 110^\circ$ denotes the orientation used in all subsequent experiments.
    }\label{fig:4ion_cyc}
\end{figure}

\begin{figure}[!b]
	\centering
    \includegraphics[width = 116 mm]{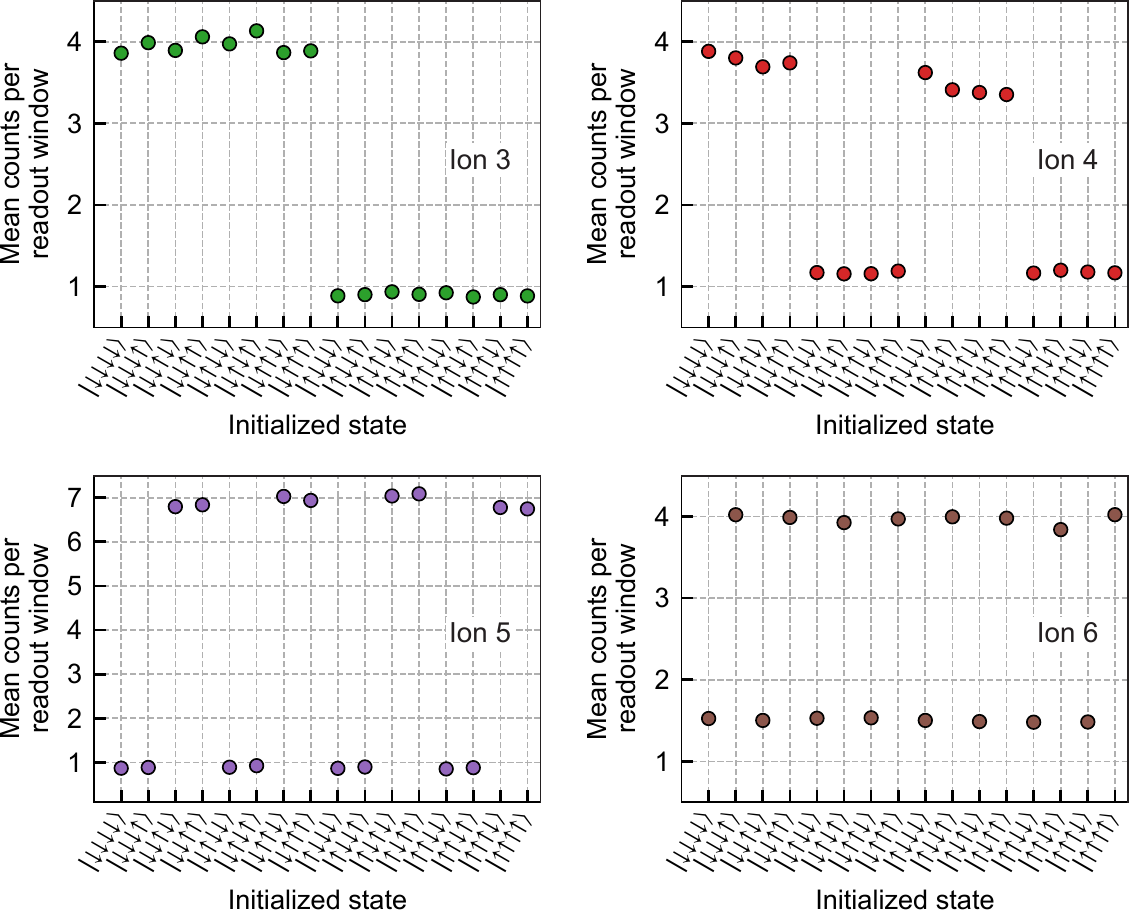}
    \caption{
    \textbf{Measured photon counts per shot for four-ions.}
    Average photon counts detected within the readout sequence
    for each ion while implementing the pulse sequence in Fig.~4a with varying initialized states but retaining the same readout sequence. 
    \ket{\downarrow\uparrow\downarrow\uparrow} corresponds to the product state $\ket{\downg}_\text{ion 3}\!\otimes\!\ket{\upg}_\text{ion 4}\!\otimes\!\ket{\downg}_\text{ion 5}\!\otimes\!\ket{\upg}_\text{ion 6}$.
    }\label{fig:CPLSimulRead}
\end{figure}

Similar to ion~1 and ion~2 in Fig.~2a, these ions also exhibit strong magnetic field orientation dependence of the spin-conserving transitions' cyclicity (Fig.~\ref{fig:4ion_cyc}). A magnetic field configuration of ($B$, $\theta$, $\varphi$) = (112 G, 90$^\circ$, 110$^\circ$) is chosen as an effective compromise that enables high cyclicities (780, 840, 750, and 850, respectively while using optical $\pi$-pulses) on the readout transitions, which are different for each ion and chosen based on their relative spectral position with respect to the cavity resonance. Note that these cyclicites are slightly different from those in Fig.~\ref{fig:4ion_cyc}, which were obtained with long, saturating optical pulses instead of $\pi$-pulses. 

We perform state-selective initialization experiments equivalent to the ones discussed in Sec.~\ref{sec:init}, achieving near-unity initialization fidelity for all four ions with 50 repetitions of the initialization sequence. 
Following optimization on the number of readout pulses for each ion, we implement the pulse sequence shown in Fig.~4a to perform simultaneous single-shot spin measurements. The average photon counts detected within the readout window for each ion is shown in Fig.~\ref{fig:CPLSimulRead} for varying combinations of four-ion initial states (note that the four-ion readout configuration remains same throughout the experiment). For each ion, the detected photon counts are observed to be dependent only on its spin state and seemingly uncorrelated to operations performed on the other ions, indicating that measurement crosstalk in this multi-qubit initialization and readout scheme is small. 

\bibliography{phase_gate_SI.bib}